\documentclass[twocolumn]{aa} 
\usepackage{natbib}
\usepackage{amsmath}
\usepackage{graphicx}
\usepackage{txfonts}
\usepackage{textcomp}
\usepackage{booktabs}
\usepackage{calligra}
\usepackage[T1]{fontenc}
\makeatletter
\def\@biblabel#1{}
\makeatother

\newcommand{\mjup}{M$_{\rm {\tiny jup}}$\,}
\newcommand{\msun}{M$_\odot$\,}
\newcommand{\mstar}{M$_\star$}
\newcommand{\ms}{m\,s$^{-1}$\,}
\newcommand{\teff}{T$_{{\rm eff}}$\,}
\newcommand{\mplan}{m$_b$\,sin$i$}
\newcommand{\Hipparcos}{{\sc Hipparcos} }
\renewcommand{\cite}{\citealp}

\begin{document}

\title{An eccentric companion at the edge of the brown dwarf desert orbiting the 2.4 \msun giant star HIP\,67537.
\thanks{Based on observations collected at La Silla - Paranal Observatory under
programs ID's 085.C-0557, 087.C.0476, 089.C-0524, 090.C-0345 and through the Chilean Telescope Time under programs
ID's CN-12A-073, CN-12B-047, CN-13A-111, CN-2013B-51, CN-2014A-52, CN-15A-48, CN-15B-25 and CN-16A-13.}}

  \titlerunning{}
   \author{M. I. Jones \inst{1,2}
           \and R. Brahm \inst{3,4}
           \and R. A. Wittenmyer\inst{5,6}
           \and H. Drass \inst{2}
           \and J. S. Jenkins \inst{7}
           \and C. H. F. Melo \inst{1}
           \and J. Vos \inst{8}
           \and P. Rojo \inst{7}}
         \institute{European Southern Observatory, Alonso de C\'ordova 3107, Vitacura, Casilla 19001, Santiago, Chile \\\email{mjones@eso.org}
         \and Center of Astro-Engineering UC, Pontificia Universidad
         Cat\'olica de Chile, Av. Vicu\~{n}a Mackenna 4860, 7820436 Macul, Santiago, Chile 
         \and Instituto de Astrof\'isica, Facultad de F\'isica, Pontificia Universidad Cat\'olica de Chile, Av. Vicu\~{n}a Mackenna
         4860, 7820436 Macul, Santiago, Chile         
         \and Millennium Institute of Astrophysics, Santiago, Chile
         \and Computational Engineering and Science Research Centre, University of Southern Queensland, Toowoomba, Queensland 4350, Australia
         \and School of Physics and Australian Centre for Astrobiology, University of New South Wales, Sydney 2052, Australia
         \and Departamento de Astronom\'ia, Universidad de Chile, Camino El Observatorio 1515, Las Condes, Santiago, Chile
         \and Instituto de F\'isica y Astronom\'ia, Universidad de Vapara\'iso, Casilla 5030, Valpara\'iso, Chile
}

   \date{}

 
\abstract{We report the discovery of a substellar companion around the giant star HIP\,67537. Based on precision radial velocity measurements
from CHIRON and FEROS high-resolution spectroscopic data, we derived the following orbital elements for HIP\,67537\,$b$: \mplan = 11.1$^{+0.4}_{-1.1}$ \mjup, 
$a$ = 4.9$^{+0.14}_{-0.13}$ AU and $e$ = 0.59$^{+0.05}_{-0.02}$. Considering random inclination angles, this object has $\gtrsim$ 65\% probability to 
be above the theoretical deuterium-burning limit, thus it is one of the few known objects in the planet to brown-dwarf transition region. 
In addition, we analyzed the {\sc Hipparcos} astrometric data of this star, from which we derived a minimum inclination angle for the companion of $\sim$ 2 deg. 
This value corresponds to an upper mass limit of $\sim$ 0.3 \msun, therefore the probability that  HIP\,67537\,$b$ is stellar in nature is $\lesssim$ 7\%.
The large mass of the host star and the high orbital eccentricity makes HIP\,67537\,$b$ a very interesting and rare substellar object.
This is the second candidate companion in the {\it brown dwarf desert} detected in the sample of intermediate-mass stars targeted by the EXPRESS radial
velocity program, which corresponds to a detection fraction of $f$ = 1.6$^{+2.0}_{-0.5}$\,\%.
This value is larger than the fraction observed in solar-type stars, providing new observational evidence of an enhanced formation efficiency of 
massive substellar companions in massive disks. 
Finally, we speculate about different formation channels for this object.}

   \keywords{techniques: radial velocities - Planet-star interactions - (stars:) brown dwarfs}

   \maketitle
%

\section{Introduction}

So far, more than 2000 exoplanets have been detected and confirmed, most of these via radial velocity 
(RV) time-series and transit observations, and thousands of new candidates from the space mission {\it Kepler} that still 
await confirmation. Soon after the discovery of the first extra-solar planets, several interesting observational results emerged, 
some of which were unexpected, showing us that planetary systems are quite common and are found to have a large diversity of orbital configurations.
In particular, the early discovery of a large population of {\it Hot-Jupiters} (including the 51 Peg system; Mayor \& Queloz \cite{MAY95}), 
the {\it planet-metallicity correlation} (Gonzalez \cite{GON97}), and the observed high eccentricity systems, among others, gave us important clues about the 
formation mechanisms and evolution of planetary systems. In addition, RV surveys have also revealed the intriguing paucity of brown dwarf (BD) companions to
solar-type stars with orbital separation $\lesssim$ 3\,-\,5 AU (Marcy \& Butler \cite{MAR00}; Marcy et al. \cite{MAR05}; Grether \& Lineweaver 
\cite{GRE06}; Sahlmann et al. \cite{SAH11}), dubbed the {\it brown dwarf desert}. \newline\indent
According to the IAU definition (Boss et al. \cite{BOS03}), a BD corresponds to a substellar object that is massive enough to burn deuterium, 
but it is not able to sustain Hydrogen fusion in its core. 
In terms of mass, these limits correspond to $\sim$\,13\,-\,80\, \mjup, for a solar composition (Chabrier \& Baraffe \cite{CHA97}; Burrows et al. \cite{BUR01}). 
Although the upper mass limit is well justified, there is no physical reason to adopt the deuterium-burning limit as a discriminant between planets and brown 
dwarfs.  Moreover, it has been argued that these types of substellar objects should be distinguished by their formation mechanism, which seems to have separate 
channels (Chabrier et al. \cite{CHA14}; Ma \& Ge \cite{MA14}). 
For instance, the fraction of giant planets (M$_{\rm p}$ $\gtrsim$ 0.5 \mjup) with $a \lesssim$ 5 AU increases from $f$ = 2.5\,$\pm\,$0.9\,\% around M dwarfs 
(Johnson et al. \cite{JOHN10}) to $f$ = 6.6\,$\pm$\,0.7\,\% for solar-type stars (Marcy et al. \cite{MAR05}; Johnson et al. \cite{JOHN10}). 
This fraction reaches a maximum value of 13.0$^{+10.1}_{-4.1}$\,\%, at $\sim$ 2 \msun (Jones et al. \cite{JON16}). Similarly, Reffert et al. (\cite{REF15})
found a peak in the detection fraction at \mstar = 1.9$^{+0.1}_{-0.5}$ \msun.
In addition, it is now well established that the fraction of giant planets around solar-type stars increases with the stellar metallicity
(Santos et al. \cite{SAN01}; Fischer \& Valenti \cite{FIS05}; Jenkins et al. \cite{JEN17}), which has been shown to be also valid for giant
(intermediate-mass) stars (Reffert et al. \cite{REF15}; Jones et al. \cite{JON16}; Wittenmyer et al. \cite{WIT17}).
These trends are in accordance with the core-accretion formation model of giant planets (Pollack et al. \cite{POL96}; Alibert et al. \cite{ALI04};
Kennedy \& Kenyon \cite{KEN08}). In contrast, BD companions are rarely found around solar-type stars interior to $\sim$ 5 AU ($f$ $\lesssim$ 0.6 \%; 
Marcy \& Butler \cite{MAR00}; Sahlmann et al. \cite{SAH11}) and also there is no clear dependence between the host star metallicity and the detection rate of 
such objects (although searching for such a correlation in the BD host stars is hampered by the very low detection rate of such objects).
In this context, it seems reasonable to believe that giant planets are efficiently formed via core-accretion in the protoplanetary disk, while BDs are 
born akin to low-mass stars, by molecular cloud fragmentation (Luhman et al. \cite{LUH07}; Joergens \cite{JOE08}), and thus 
we might expect an overlapping mass (transition) region, in which both of these formation channels take place. Therefore, the detection and characterization
of planet to BD transition objects is of key importance to better understand the thin transition regime between the high-mass planetary tail and the 
low-mass brown dwarf regime.
In particular, the mass and metallicity of the parent star certainly gives us important clues regarding the formation mechanism of such objects. \newline \indent
In this paper we present precision RVs of the intermediate-mass evolved star HIP\,67537, revealing the presence of a substellar object in the transition 
limit between giant planets and BDs. The host star is one of the targets of the {\bf EX}o{\bf P}lanets a{\bf R}ound {\bf E}volved {\bf S}tar{\bf S} 
(EXPRESS) radial velocity program (Jones et al. \cite{JON11}). Also, we analyzed the {\sc Hipparcos} astrometric data of HIP\,67537 from which we derived an upper mass limit for its companion. Finally, we discuss about the fraction of companions in the {\it brown dwarf desert} around intermediate-mass stars and we speculate on the different  scenarios that might explain the formation and orbital evolution of this system. 
The paper is organized as follows: In section 2 the observations, data reduction and orbital solution are presented. In section 3 we present 
in detail our new codes that we use to compute the radial velocities, for both the simultaneous calibration method and the I$_2$ cell technique.
In section 4 we present the physical properties of HIP\,67537, while its companion orbital elements are presented in section 5. In section 6, we present a 
detailed study of the photometric variability, bisector analysis and chromospheric stellar activity of the 
host stars. In section 7 we analyze the {\sc Hipparcos} astrometric data of HIP\,67537 and its companion upper mass limit. Finally, the summary and discussion is 
presented in section 8.

\section{Observations and data reduction}

The observations were performed with the FEROS (Kaufer et al.  \cite{KAU99}) and CHIRON (Tokovinin et al. \cite{TOK13}) high-resolution optical spectrographs.
FEROS is equipped with two fibres, one for the science object and the second one for simultaneous calibration, which is used to track and correct the 
spectral drift during the observations (see Baranne et al. \cite{BAR96}). 
The reduction of the FEROS data was done in the standard fashion (i.e, bias subtraction, flat-field 
correction, order-by-order extraction and wavelength calibration) using the CERES reduction code (Jord\'an et al. \cite{JOR14}; Brahm et al. \cite{BRA16}). 
On the other hand, CHIRON is equipped with an iodine cell, which is located in the light path, in front
of the fibre entrance in the spectrograph. The I$_2$ vapor inside the cell absorbs part of the incoming light, producing a rich narrow absorption spectrum that
is superimposed onto the stellar spectrum, in the range between $\sim$ 5000-6200 \AA. We use the CHIRON pipeline to obtain order-by-order wavelength
calibrated spectra. We typically use the fiber slicer, which delivers a spectral resolution of $\sim$ 80,000, and much higher efficiency compared to the {\it slit} 
(R$\sim$ 90,000) and {\it narrow slit} mode (R$\sim$ 130,000).

\section{Radial velocities}

We have recently developed new radial velocity analysis codes for both FEROS and CHIRON data.
In the two cases, we have reduced our internal RV uncertainties by up to a factor two.
Additionally, we have developed automatic stellar activity diagnoses that are included in these new pipelines.
The new main features and differences with the old codes (e.g. Jones et al. \cite{JON13}; Jones \& Jenkins \cite{JON14}) are discussed in the following sections.

\subsection{FEROS data \label{sec_FEROS}}
\begin{figure}[]
\centering
\includegraphics[width=6cm,height=7.5cm,angle=270]{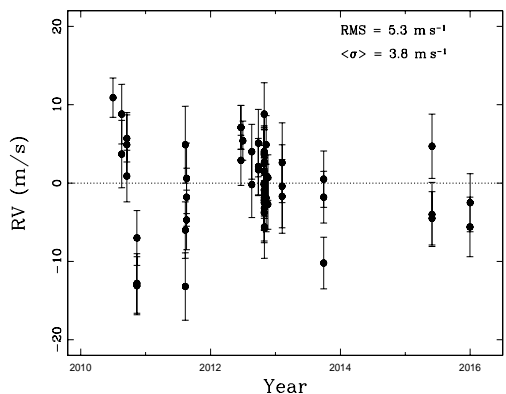}
\caption{5.5 years of FEROS observations of the RV standard star $\tau$ Ceti. The mean internal error is 3.8 \ms, while the long-term RMS around the mean is 5.3 \ms.
\label{tau_ceti}}
\end{figure}

The FEROS radial velocity variations were computed using the cross-correlation technique (Tonry \& Davis \cite{TON79}), with a new dedicated IDL-based pipeline, 
which is more flexible and user-friendly that our old IRAF and Fortran based codes used for this purpose (Jones et al. \cite{JON13}). 
We compute the cross-correlation function (CCF) between a high S/N template, which is created by stacking all of the FEROS spectra of each star, 
after correcting by their relative velocity offset, and each observed spectrum. We then fit the CCF by a Gaussian plus a linear function. 
We note that the addition of the linear term improves our results when compared to the single Gaussian CCF model.
The maximum of the fit corresponds to the wavelength (velocity) shift. This method is applied to a total of 100 chunks per spectrum, each of $\sim$ 50\,\AA\, 
in length, across 25 different orders, covering the wavelength range of $\sim$ 3900\,-\,6700 \AA. 
Then, deviant chunk velocities are filtered-out using a 3-$\sigma$ iterative rejection method. The velocity shift per epoch is computed
from the median of the non-rejected chunk velocities and its uncertainty corresponds to the formal error in the mean\footnote{$\sigma_{RV} = \sqrt{\sigma_c}/(n-1)$, where $n$ is the number of non-rejected chunks and $\sigma_c$ corresponds to the RMS of the non-rejected 
velocities.}. We note that we use the median instead of the mean because it leads to slightly better results in terms of long-term stability observed in the RV 
standard star $\tau$ Ceti. 
A similar procedure is computed for the simultaneous calibration lamp. However, in this case the template 
corresponds to the lamp observation that is used to compute the wavelength solution. The final velocities are obtained after correcting the night drift recorded
by the simultaneous lamp and the barycentric correction, which is computed at the mid-time of the observation (FEROS is not equipped with an exposure
meter).  We note that we assign a constant weight to all of the non-rejected chunks. We tried different weighting scenarios based on different combinations of the CCF parameters (height and width), but no improvement in the final velocities was observed.
Figure \ref{tau_ceti} shows 57 FEROS RV epochs spanning a total of 5.5 years of the standard RV star $\tau$ Ceti. 
The mean internal uncertainty is 3.8 \ms. The long-term stability is 5.3 \ms, which is superior to the value of $\sim$ 10 \ms (restricting to the observations 
taken after 2010) obtained with the ESO data reduction system for FEROS (Soto et al. \cite{SOT15}).

\subsection{CHIRON data}
The CHIRON velocity variations were computed using a similar method as presented in Butler et al. (\cite{BUT96}), however we use a simpler PSF model, including only 
one Gaussian (being the width of the Gaussian a free parameter), which yields nearly identical results to the multi-Gaussian models. 
Also, we compute the radial velocities for a total of 352 chunks, each of 180 pixels, spread over 22 
different orders. The resulting velocity at each epoch is obtained from the median in the individual chunks velocities, after passing an iterative 
rejection procedure, in a similar fashion as done for the FEROS data. 
The typical RV precision that we achieve is $\sim$ 3 \ms for {\it slit} observations (R $\sim$ 90,000) and $\sim$ 4 \ms using the {\it image slicer} (
R $\sim$ 80,000). We note that it is possible to achieve a precision $\sim$ 2 \ms applying the single-Gaussian model to high signal-to-noise (S/N) ratio observations using the narrow-slit mode (R$\sim$ 130,000), but at a cost of much higher exposure 
times due to the reduced efficiency. In particular, the RV precision is highly dependent on the quality of the stellar
template, which is constructed via PSF deconvolution of a I$_2$-free observation of the star. However, due to the intrinsic p-modes induced RV variability of all
of our targets (typically at the $\sim$ 5-10 \ms level; see Kjeldsen \& Bedding \cite{KJE95}), we have adopted the {\it image slicer} mode,
which provides higher throughput compared to the slit modes and allow us to achieve instrumental uncertainties below the stellar noise level.

\section{HIP\,67537 properties}

The fundamental parameters of HIP\,67537 are listed in Table \ref{orb_par}. The visual magnitude, B-V color, and the corresponding errors were computed
from the linear transformations between the Tycho and Johnson photometric systems, as given in section 1.3 of the {\sc Hipparcos} and Tycho Catalogs 
(ESA, \cite{ESA97}). The distance to the star was computed using the parallax listed in the new data reduction of the {\sc Hipparcos} data (Van Leeuwen, \cite{VAN07}).
We note that no parallax for HIP\,67537 is available from the GAIA DR1.
We corrected the visual magnitude using the Arenou et al. (\cite{ARE92}) extinction maps and applied the Alonso et al. (\cite{ALO99}) bolometric correction to 
obtain the stellar luminosity.
The atmospheric parameters, namely T$_{\rm eff}$, log{\it g} and [Fe/H], were derived using the equivalent width ($W$) of a carefully selected list of $\sim$ 
150 Fe\,{\sc i} and $\sim$ 20 Fe\,{\sc ii} relatively weak lines ($W\,\lesssim$\,150\,\AA\,), which were measured using the 
ARES\footnote{http://www.astro.up.pt/$\sim$sousasag/ares/} code (Sousa et al. \cite{SOU07}). For this purpose, we used 
MOOG\footnote{http://www.as.utexas.edu/$\sim$chris/moog.html} (Sneden \cite{SNE73}), 
which solves the radiative transfer equation, imposing local excitation and ionization equilibrium. 
Briefly, for a given set of atmospheric parameters, MOOG computes the iron abundance corresponding to each measured equivalent width, by matching the curve 
of growth in the weak line regime, and including the effect of the micro-turbulence. The final atmospheric parameters 
are thus obtained in an iterative process, by removing any dependence between the abundance with the excitation potential and reduced equivalent widths ($W/\lambda$), and 
also by forcing the iron abundance to be the same from both species (Fe\,{\sc i} and Fe\,{\sc ii}). For a detailed description of this method see 
(Gray \cite{GRAY05}). The resulting atmospheric parameters of HIP\,67537 are listed in Table \ref{orb_par}. For
comparison, Alves et al. (\cite{ALV15}), based o a similar approach, obtained the following parameters: \teff = 5017 $\pm$ 042 K,
log$g$ = 3.08 $\pm$ 0.08 cm\,s$^{-2}$ and [Fe/H] = 0.17 $\pm$ 0.03 dex. These results are in good agreement with those presented here.
Finally, the stellar position in the H-R diagram and the derived metallicity were compared with Salasnich et al. (\cite{SAL00})
evolutionary tracks, to obtain the stellar mass and radius. This procedure was repeated 100 times, from random generated datasets, assuming Gaussian distributed
errors in the luminosity, effective temperature and stellar metallicity. The adopted values for \mstar\, and R$_\star$, and their corresponding
uncertainties, were obtained from the mean and standard deviation in the resulting distribution from the 100 random samples. For further details 
see Jones et al. (\cite{JON11,JON15b}) 

\section {Orbital elements of HIP\,67537\,$b$}

\begin{figure}[]
\centering
\includegraphics[width=8cm,height=9.5cm,angle=270]{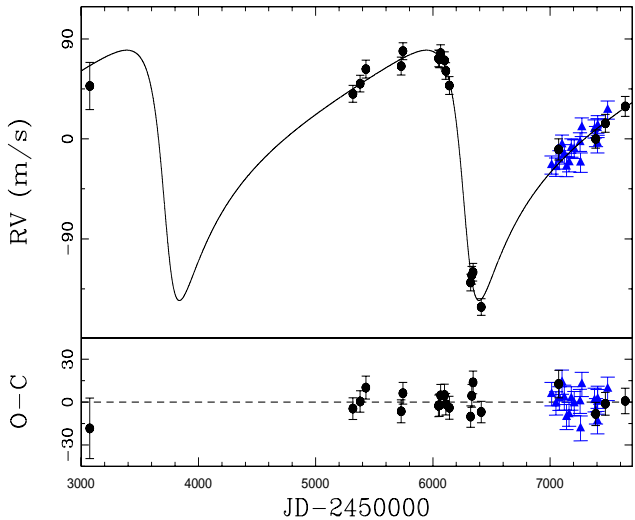}
\caption{Radial velocity measurements of HIP\,67537. The black circles and blue triangles represent the FEROS and CHIRON velocities, 
respectively. The best Keplerian solution is overplotted (black solid line). The post-fit residuals are shown in the lower panel. 
\label{fig_HIP67537_vels}}
\end{figure}

\begin{figure}[]
\centering
\includegraphics[width=6cm,height=8.5cm,angle=270]{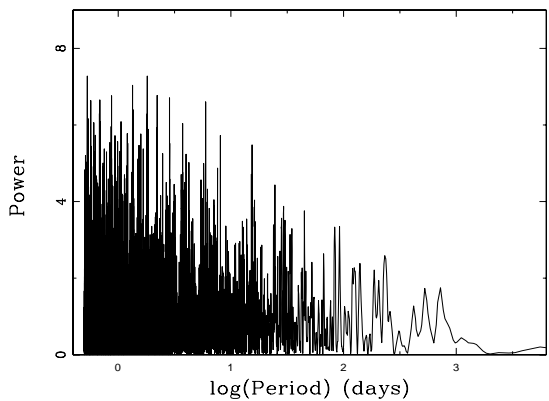}
\caption{Lomb-Scargle periodogram of the post-fit residuals of the HIP\,67537 velocities. \label{fig_residuals}}
\end{figure}

We obtained a total of 19 FEROS spectra and 18 CHIRON observations of HIP\,67537, covering a total baseline of more than 6 years.
In addition, we retrieved a FEROS observation from the ESO archive, which was taken in 2004, but without simultaneous calibration.
However, since FEROS is relatively stable\footnote{FEROS is thermally stabilized, with temperature variations typically $\lesssim$ 0.15 K.} 
the spectral drift was computed from three RV stable stars that were 
observed before and after HIP\,67537. The night drift was then interpolated to the time of the observation of HIP\,67537. 
We note that we have applied this method to FEROS data of HIP\,67851 (which were taken immediately after HIP\,67537) to constrain the 
orbital period of HIP\,67851\,$c$ (Jones et al. \cite{JON15b}). New RV measurements of HIP\,67851 (which already cover one 
orbital period of HIP\,67851\,$c$) confirm the validity of this method. 
The resulting radial velocities are listed in Table \ref{orb_par} and are shown in Figure \ref{fig_HIP67537_vels}. 
As can be seen, the peak-to-peak variation exceeds 200 \ms, which is 
indicative of the presence of a massive substellar object. The orbital elements of the companion were obtained with the Systemic Console\footnote{http://oklo.org} 
(Meschiari et al. \cite{MES09}), after adding 7 \ms RV noise in quadrature to the radial velocities, which is the typical level of RV scatter observed in our sample.
The resulting values are listed in Table \ref{orb_par}. The uncertainties were derived using the bootstrap tool included in version 
2.17 of Systemic and correspond to the 1-$\sigma$ equal-tailed confidence interval.
The best Keplerian fit is overplotted in Figure \ref{fig_HIP67537_vels}. The RMS about the best fit is 8.0 \ms. We note that no significant 
improvement in the Keplerian fit is obtained by including a linear trend in the solution and no significant periodicity is present in the post-fit residuals
(see Figure \ref{fig_residuals}).

\begin{table}
\centering
\caption{Stellar properties and orbital elements.\label{orb_par}}
\begin{tabular}{lr}
\hline\hline
\vspace{-0.3cm} \\
Stellar properties of HIP\,67537& \\
\hline \vspace{-0.3cm} \\
Spectral Type         &        K1III       \\                                                                
$B-V$ (mag)           &  0.99  $\pm$ 0.006 \\                                                                
$V$ (mag)             &  6.44  $\pm$ 0.005 \\
Distance (pc)         &  112.6 $\pm$ 5.8   \\
\teff (K)             &  4985  $\pm$  100  \\
Luminosity (L$_\odot$)         &  41.37 $\pm$  7.16 \\
log\,$g$ (cm\,s$^{-2}$) &  2.85  $\pm$  0.2  \\
{\rm [Fe/H]} (dex)    &  0.15  $\pm$  0.08 \\
$v$\,sin$i$ (k\ms)    &  2.3  $\pm$  0.9  \\
M$_\star$ (\msun)     &  2.41  $\pm$  0.16 \\
R$_\star$ (R$_\odot$) &  8.69  $\pm$  0.88 \\
\vspace{-0.3cm} \\
\hline \vspace{-0.1cm} \\
 Orbital parameters of HIP\,67537\,{\it b}  &              \\  
\hline \vspace{-0.3cm}                                     \\ \vspace{0.2cm}
P (days)                     &   2556.5$^{+99.2}_{-94.9}$  \\ \vspace{0.2cm}
K (\ms)                      &  112.7$^{+7.8}_{-3.0}$       \\ \vspace{0.2cm}
$a$ (AU)                     &   4.91$^{+0.14}_{-0.13}$     \\ \vspace{0.2cm}
$e$                          &   0.59$^{+0.05}_{-0.02}$    \\ \vspace{0.2cm}
\mplan\, (\mjup)             &   11.1$^{+0.4}_{-1.1}$      \\ \vspace{0.2cm}
$\omega$ (deg)               &  119.6$^{+4.6}_{-6.7}$      \\ \vspace{0.2cm}
T$_{\rm P}$ (JD-2450000)     & 6290.6$^{+16.2}_{-51.6}$    \\ \vspace{0.2cm}
$\gamma_1$ (\ms) (FEROS)     &   11.0$^{+3.8 }_{-3.8 }$    \\ \vspace{0.2cm}
$\gamma_2$ (\ms) (CHIRON)    &   -6.1$^{+3.7 }_{-3.7 }$    \\ \vspace{0.2cm}
RMS (\ms)                    &  8.0                \\  
$\chi$$^2_{\rm red}$         &  1.0                \\  
\vspace{-0.3cm} \\\hline\hline
\end{tabular}
\end{table}

\section{Planet validation}

Stellar phenomena such as non-radial pulsations, spots and plages in rotating stars and other activity-related effects (like suppression of 
the convective blueshift in active regions), might produce apparent RV variations, mainly via CCF deformation, that can mimic the effect of a genuine 
doppler signal induced by an orbiting companion (e.g. Saar \& Donahue \cite{SAR97}; Hu\'elamo et al. \cite{HUE08}; Meunier et al. \cite{MEU10}; 
Dumusque et al. \cite{DUM11}).
In the following sections we analyze the available {\sc Hipparcos} photometric data, we present a study of the CCF asymmetry variations and a chromospheric 
activity analysis, to understand whether the RV variations observed in HIP\,67537 are explained by intrinsic stellar phenomena, like 
those discussed above.

\subsection{Photometric analysis \label{phot_analisys}}

We analyzed the {\sc Hipparcos} photometry of HIP\,67537, to investigate a possible correlation with the radial velocities.
This dataset consists of a total of 94 good quality measurements (quality flag equal to 0 and 1), covering a time span of 1164 days, which is 
is significantly shorter than the orbital period, thus it is not possible to search for periodic photometric signals with similar periods than the orbital one.
However, the data present a variability of only 0.006 mag (corresponding to $\sim$ 0.6\% in flux), which is too small to explain the large velocity 
variations observed in slow rotating stars like HIP\,67537 (Hatzes \cite{HAT02}; Boisse et al. \cite{BOI12}). 
Moreover, in this scenario we would expect the radial velocity period to match the stellar rotational 
period, which is clearly not the case. Based on the measured $v$\,sin\,$i$ and R$_\star$ (see Table \ref{orb_par}),
we expect a maximum stellar rotational period of $\sim$ 191 days, which is $\sim$ 15 times shorter than the observed orbital period. 
Therefore, we discard rotational modulation as the cause of the observed RV variations. \newline \indent
Additionally, to test for possible light contribution from the unseen companions, we used Johnson, GENEVA and 2MASS photometric data from
the literature. 
The fitting procedure used is the binary SED fit outlined in Vos et al. (\cite{VOS12,VOS13}), in which the parameters of the giant component are kept
fixed (to those listed in Table \ref{orb_par}), while companion parameters are varied. Furthermore, the {\sc Hipparcos} parallax is
used as an extra constraint. For this procedure, five photometric points are enough for a reliable result (e.g. Bluhm et al. \cite{BLU16}).
The observed photometry is fitted with a synthetic SED integrated from the Kurucz (\cite{KUR79}) atmosphere models ranging in effective temperature from 3000
to 7000 K, and in surface gravity from $\log{g}$=2.0 dex (cgs) to 5.0 dex (cgs). The radius of the companion is varied from 0.1 R$_{\odot}$ to 2.0 $R_{\odot}$. 
The SED fitting procedure uses the grid based approach described in Degroote et al. (\cite{DEG11}), where 10$^6$ models are randomly picked in
the available parameter space. The best fitting model is determined based on the $\chi^2$ value.
As the parameters (effective temperature, surface gravity and radius) of the giant component are fixed at the values determined from the spectroscopy, and the
distance to these systems is known accurately from the {\sc Hipparcos} parallax, the total luminosity of the giant is fixed. This allows to accurately determine the
amount of extra light from the companion, based on the SED fit.
For this system, this is less than 1\%, which is within the uncertainties of the SED fit. We can thus conclude that 
no significant light contribution from the companion is observed in this system. 
As an extra test, an unconstrained SED fit was performed, in which the atmospheric parameters of the giant were varied.
This provides an independent set of atmospheric parameters. We find that in both cases the atmospheric parameters of the best fitting SED models correspond well
with those derived from spectroscopy. We found no indication of contamination from an unseen companion. 

\subsection{Line asymmetry}

We computed the bisector velocity span (BVS) of the CCF (Toner \& Gray \cite{TON88}), as a stellar line asymmetry indicator, since spots in a rotating star and 
non-radial pulsations propagating in the stellar surface can produce significant distortions in the observed stellar spectral lines.
For FEROS spectra, we computed the BVS of the CCF, for each of the 100 chunks (see section \ref{sec_FEROS}). Similarly to the RV values, the resulting BVS value 
at each epoch is obtained from the mean in the 100 BVS values, after passing a 3-$\sigma$ iterative rejection method. The corresponding uncertainty is 
derived simply as the error in the mean of the non-rejected BVS values. Similarly, we computed the full-width-at-half-maximum (FWHM) of the CCF at each epoch. 
\newline \indent
In the case of CHIRON data, we cannot apply the same method, since the RV are not computed via cross-correlation and also because the spectra are 
contaminated by the I$_2$ cell absorption spectrum in the wavelength range of $\sim$ 5000\,-\,6200\,\AA\,. However, we take advantage of the fact that there are 
still many I$_2$-free orders, that are useful to measure variations in the stellar absorption lines profile. Essentially, we use the CHIRON 
I$_2$-free wavelength range, which corresponds to 36 orders covering between $\sim$ 4600\,-\,5000\,\AA\, and $\sim$ 6250\,-\,8750\,\AA\,. We then computed the CCF 
between each template and the observations, in exactly the same manner as done for FEROS spectra, as described in $\S$ \ref{sec_FEROS}, but this time using only 
two chunks per order, which we found leads to the smaller uncertainties in both, the RV and BVS values. The corresponding uncertainties are computed as for the 
FEROS CCF, as explained above. The resulting BVS and FWHM variations versus the RVs are displayed in Figure \ref{HIP67537_act} (upper and middle panel, respectively). 
As can be seen, although there is some level of correlation between the FEROS RVs and BVS, it is mainly explained by the three datapoints around $\sim$ -100 \ms. 
In fact, other stars that we observed during those three nights also present BVS significantly higher than their mean value. 
We thus conclude that this observed relationship is mainly explained by an instrumental effect (instrumental profile variations, poor fibre scrambling,
etc) rather than an intrinsic stellar effect. On the other hand, despite one outlier that is above the mean, the FWHM variations show no significant correlation with 
the RVs.

\subsection{Chromospheric activity}

\begin{figure}[]
\centering
\includegraphics[width=8cm,height=9.5cm,angle=270]{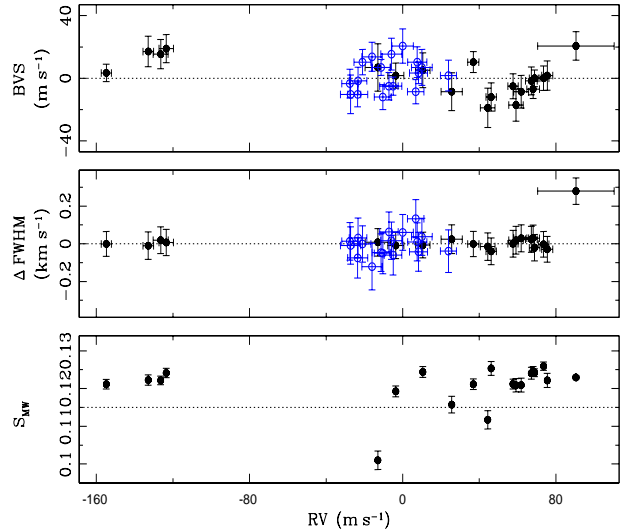}
\caption{BVS, FWHM and S-index variations versus the FEROS (black filled circles) and CHIRON (blue open circles) velocities for HIP\,67537 
({\it upper, middle} and {\it lower panel}, respectively.) 
\label{HIP67537_act}}
\end{figure}

We computed the activity S-index variations from the chromospheric re-emission in the core of the Ca\,{\sc ii} H ($\lambda$ = 3933.67\AA) and Ca\,{\sc ii} 
K ($\lambda$ = 3968.47\AA) lines. 
For this purpose, we measured the S-indexes from FEROS spectra (CHIRON does not reach this wavelength regime) in a similar fashion as described in Jenkins et al. 
(\cite{JEN08}). We calibrated our FEROS S-indexes to the Mount Wilson system (MWS), using 10 stars listed in Duncan  et al. (\cite{DUN91}). We apply a simple linear 
correlation between the FEROS system and the MWS (e.g. Tinney et al. \cite{TIN02}; Jenkins et al. \cite{JEN06}). 
The uncertainties correspond to the error in the S-index, which is due to photon noise statistics. 
The lower panel in Figure \ref{HIP67537_act} shows the resulting S-values in the MWS (S$_{\rm MW}$), versus the FEROS 
radial velocities. 
Clearly, there is no dependence between the S$_{\rm MW}$ indexes and the RVs. 
Based on these results, and due to the long orbital period observed, we discard that spots, activity or stellar pulsations as the cause of the observed RV 
variations, confirming the planetary hypothesis. 

\section{Astrometric upper mass limit \label{astrometry}}

\begin{figure}[]
\centering
\includegraphics[width=8cm,height=9.5cm,angle=270]{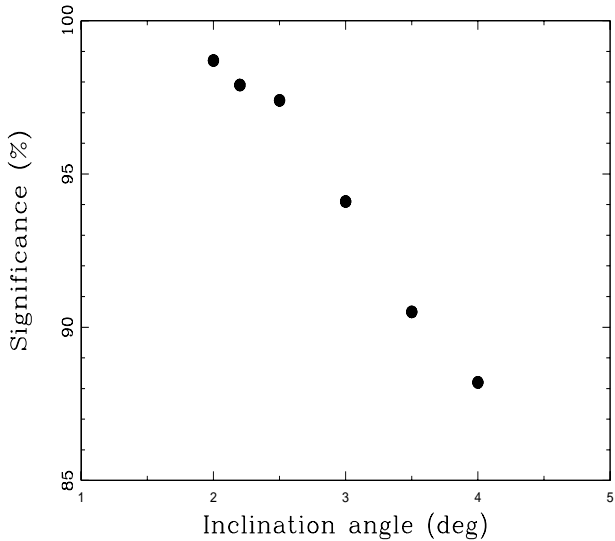}
\caption{Significance of the solution to the synthetic datasets as a function of the inclination angle.
\label{HIP67537_inc_significance}}
\end{figure}

Motivated by previous works (Reffert \& Quirrenbach \cite{REF11}; Sahlmann et al. \cite{SAH11}; D\'iaz et al. \cite{DIA12}), we used the improved version of the 
\Hipparcos astrometric data (Van Leeuwen \cite{VAN07}), to measure the inclination angle of the orbital plane, and thus to derive the actual mass of HIP\,67537\,$b$. 
To do this, we employed the method described in Sahlmann et al. (\cite{SAH11}).
Briefly, from the residuals of the \Hipparcos abscissa, we reconstructed the abscissa values, and 
recomputed the astrometric solution, but this time solving for 7-parameters, i.e., the parallax ($\varpi$), celestial position ($\alpha^{\star}$, $\delta$), proper 
motion ($\mu_{\alpha^{\star}}$, $\mu_{\delta}$), the inclination angle ($i$) and the longitude of the ascending node ($\Omega$). 
Using this method, several exoplanet and BD candidates have recently been confirmed (e.g. Wilson et al. \cite{WIL16}). 
We note that we have tested our method using some of the systems presented in Sahlmann et al. (\cite{SAH11}) and Wilson et al. (\cite{WIL16}), for which we 
obtained nearly identical results. An extensive description of the method and validation on real data will be presented soon (Jones in preparation). \newline \indent
The {\sc Hipparcos} astrometric dataset of HIP\,67537 is comprised by a total of 105 measurements, after removing one outlier (at the $>$ 4 $\sigma$ level), with a 
mean uncertainty of 1.97 mas and covering 1164 days. 
The solution type is 5, meaning that no indication of significant acceleration in the proper motion is observed.
Unfortunately, due to the low astrometric amplitude of the signal ($a\,$sin\,$i$ = 0.19 mas) and the long orbital period of HIP\,67537\,$b$, which exceeds by a 
factor of two the \Hipparcos data timespan, no astrometric signal was detected. However, we can put an upper mass limit, corresponding to the 
minimum inclination angle that would be detectable in the \Hipparcos data. To do this, we generated synthetic astrometric datasets, including the 
gravitational effect from the unseen companion, for a given inclination angle. We note that the smaller the inclination angle, the larger the astrometric 
signal, thus the easier its detection. We then used the same method described above, but this time using the synthetic datasets, instead of the original 
\Hipparcos data. 
For each realization, we used the Keplerian parameters from the 1000 bootstrap Keplerian solutions.
We also generated Gaussian distributed errors for the \Hipparcos abscissa residuals and the $\Omega$ values were randomly choose.
Then, for each solution, we applied the permutation test, in which the dates of the \Hipparcos observations are fixed, while the corresponding abscissa residuals
are randomly permuted. The significance of the solution is set by the fraction of the permuted solutions that yield $\chi^2$ values greater 
than the original solution.
Figure \ref{HIP67537_inc_significance} shows the significance of the synthetic orbit as a function of the inclination angle. 
It can be seen that the significance of the solution increases steeply with decreasing inclination angle. For this star,
a significance of 98.7\% is reached at $i$ = 2 deg, corresponding to a maximum mass for the companion of 0.33 \msun, while it drops to $\sim$ 90\% at 
$i \sim$ 3.5 deg. 
By assuming $i$ = 2 deg as the minimum inclination angle, the probability that HIP\,67537\,$b$ is actually a stellar object is $\lesssim$ 7 \% (corresponding to
2 deg $\lesssim$ $i$ $\lesssim$ 8 deg).

\section{Summary and discussion}

\begin{figure}[]
\centering
\includegraphics[width=8cm,height=9.0cm,angle=270]{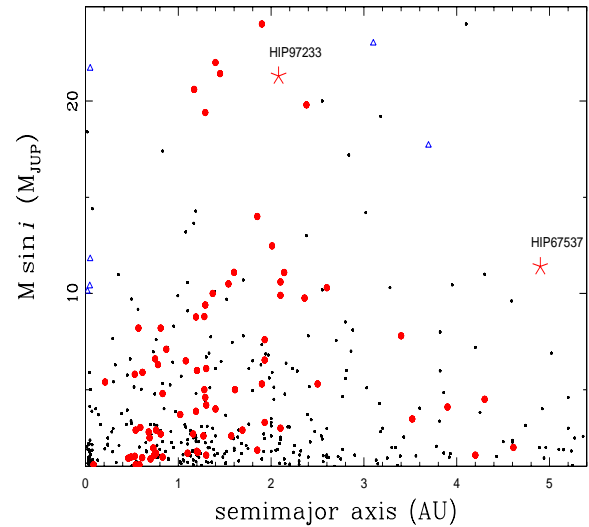}
\caption{Minimum planet mass versus semimajor axis, for known giant planets (\mplan $\gtrsim$ 1.0 \mjup). The black dots and
red filled circles correspond to main-sequence and giant host stars. The blue open triangles are those systems with known inclination angles, thus they 
correspond to the true mass of the companion. The red asterisks show the position of HIP\,67537\,$b$ and HIP\,97233\,$b$.
\label{fig_new_semiaxis_star}}
\end{figure}

In this work we present more than 6 years radial velocity variations of the evolved star HIP\,67537. 
Based on the Keplerian fit to the observed RVs and also from the astrometric orbital inclination constraints presented in the previous sections,
HIP\,67537\,$b$ is most likely a massive substellar companion, in the super-planet to brown dwarf mass regime. 
By assuming random orbital inclination angles and based in the upper mass limit of $\sim$ 0.3 \msun (at the $\sim$ 99\% significance level)
, the probability that the HIP\,67537 companion is stellar in nature is only $\sim$ 7\%.
Moreover, out of the 24 binary companions detected in our sample, only 6 of them are found interior to 5 AU.
Interestingly, these 6 companions have minimum masses $\gtrsim$ 0.3 \msun (see Bluhm et al. \cite{BLU16}). A similar result is also observed in solar-type binaries
(Raghavan et al. \cite{RAG10}). In fact, hydrodynamical simulations show that close binary systems ($a \lesssim$ 10 AU) preferentially form with mass ratios
close to unity (Bate et al. 2002).
This means that very low-mass binary companions are rarely found in relatively close-in orbits, consistent with the substellar companion hypothesis. \newline \indent 
Figure \ref{fig_new_semiaxis_star} shows the position of HIP\,67537\,$b$ in the semimajor axis versus minimum mass diagram. 
The red circles correspond to giant host stars, while the small black dots are solar-type parent stars\footnote{source: http://exoplanets.eu/}.
Clearly HIP\,67537\,$b$ is placed in a barely populated region of this diagram. In fact, apart from $\nu$ Oph\,$c$ (Quirrenbach et al. \cite{QUI11}; Sato et al. 
\cite{SAT12}), this is the only known super-planet/BD candidate known to orbit a giant star at such large orbital distance. 
Given its projected mass and semimajor 
axis, this object is located in the edge of the {\it BD desert}, making HIP\,67537\,$b$ a rare object. 
After HIP\,97233\,{\it b} (Jones et al. \cite{JON15a}), this is the second BD candidate detected by our program orbiting interior to 5 AU. 
Considering two BDs in our sample comprised by 166 stars, we obtain a fraction\footnote{Corresponding to a 68\% equal-tailed interval. See Cameron \cite{CAM11}} 
of $f$ = 1.2$^{+1.5}_{-0.4}$\,\%, higher than $f 
\sim$ 0.5\,-\,0.8\% reported by other RV surveys targeting solar-type stars (Marcy \& Butler \cite{MAR00}; Vogt et al. \cite{VOG02}; 
Wittenmyer et al. \cite{WIT09}; Sahlmann et al. \cite{SAH11}).  Interestingly, both stars have masses $\gtrsim$ 1.9 \msun, providing further indications that 
BDs are more efficiently formed around more massive stars (Lovis \& Mayor \cite{LOV07}; Mitchell et al. \cite{MIT13}), which are formed in denser environments 
and thus have more massive protoplanetary disks (Andrews et al. \cite{AND13}).
Moreover, if we restrict our sample to intermediate-mass stars (\mstar $\gtrsim$ 1.5 \msun), then the fraction of BD companions with $a \lesssim$ 5 AU rises to 
$f$ = 1.6$^{+2.0}_{-0.5}$\,\%. For comparison, Borgniet et al. (\cite{BOR16}) found no BD with orbital periods less than 1000 days, from a sample of 51 
intermediate-mass A-F dwarf stars, which are the main-sequence progenitors of GK giants (although there is a big debate on this subject; see Johnson \& Wright 
\cite{JOHN13} and references therein). This result is in agreement with our findings, since our two BD candidates have $P >$ 1000 days. \newline \indent
Interestingly, the parent stars of these BD candidates are metal-rich, therefore it is plausible that they formed via core-accretion.
According to Mordasini et al. (\cite{MOR09}) planets in massive and metal-rich disks can be formed at starting position $\sim$ 4-7 AU and can accrete a 
significant amount of mass in-situ, becoming super-planets (or BDs) prior to the disk dissipation and opening a gap in the disk. Subsequently, they move 
inward via type II migration (Papaloizou \& Lin \cite{PAP84}) to their final position at $a \gtrsim$ 2 AU. 
In addition, these two systems present high orbital eccentricities ($e$ $\sim$ 0.6), in contrast to most giant planets orbiting giant stars, which are typically found in nearly circular orbits ($e \lesssim$ 0.2; e.g. Jones et al. \cite{JON14b}). In fact, these are the only substellar objects in our sample with
eccentricities exceeding $\sim$ 0.2 (updated orbital solutions and new EXPRESS systems will be presented in a forthcoming paper). Ribas \& 
Miralda-Escud\'e (\cite{RIB07}) studied the eccentricity distribution of planets detected via RVs around solar-type stars and they found that the most
massive planets (M$_{\rm p}$ $\gtrsim$ 4 \mjup) tend to have larger orbital eccentricities than less massive objects. This observational trend has been more recently 
confirmed by Desidera et al. (\cite{DES12}) and Adibekyan et al. (\cite{ADI13}). From a theoretical point of view, the high eccentricity observed in giant planets 
can be explained by planet-planet encounters, leading to eccentricity excitation and radial migration (e.g. Rasio \& Ford \cite{RAS96}; Raymond et al. \cite{RAY10}).
Moreover, according to Ida et al. (\cite{IDA13}), massive giant planets could be formed in multi-planet systems in massive and metal-rich disks, with circular 
orbits, and due to the interaction with other planets in the system their eccentricities are excited. As a consequence, during these encounters, the less
massive planets are either ejected or scattered to wider orbits ($\gtrsim$ 30 AU). In fact, multi-planet systems comprised by two or more
giant planets are common among intermediate-mass giant stars (Jones et al. \cite{JON16}), while systems comprised by a BD and a giant planet appear to be absent. 
This could be the result of the ejection of a smaller giant planet by a BD in the system, like  HIP\,67537\,$b$ and HIP\,97233\,$b$.
The detection of outer giant planet companions using direct imaging might provide strong observational evidence of this scenario.
Other mechanisms could be also responsible for the observed high eccentricities of these systems. For instance, 
the eccentricity of super planets and BDs can be excited by a distant companion, via the Kozai-Lidov effect (Kozai \cite{KOZ62}; Lidov \cite{LID62}; 
Holman et al. \cite{HOL97}). This mechanism probably affects many planetary systems, 
given the large fraction of stellar companions observed at different stellar mass, including intermediate-mass evolved 
stars (e.g. Bluhm et al. \cite{BLU16}; Wittenmyer et al. \cite{WIT17}). Unfortunately, due to the very limited number of known close-in brown dwarf companions 
it is still very difficult to either favor or discard different formation and evolution models. The discovery of more of these systems are mandatory to really
understand how these very massive planets form and how they interact with the disk and the rest of the bodies in it, as well as to study the formation efficiency as a function of the stellar mass. 

\begin{acknowledgements}
M.J. acknowledges financial support from Fondecyt project \#3140607 and FONDEF project CA13I10203.
J.J. acknowledges funding by the CATA-Basal grant (PB06, Conicyt).
We acknowledge the referee Johannes Sahlmann for very useful comments on this article.
This research has made use of the SIMBAD database and the VizieR catalogue access tool, operated at CDS, Strasbourg, France. 
\end{acknowledgements}


\begin{appendix} 

\section{Radial velocity tables.}


\begin{table}
\centering
\caption{Radial velocity variations of HIP\,67537\label{HIP95124_vels}}
\begin{tabular}{lccc}
\hline\hline
JD\,-\,2450000        & RV & error  & Instrument\\
          &  (\ms)   &   (\ms) \\
\hline \vspace{-0.3cm} \\
3072.8542  &   36.6  &20.0& FEROS \\
5317.6184  &   29.5  &2.9 & FEROS \\
5379.6473  &   38.7  &2.7 & FEROS \\
5428.5237  &   51.8  &3.8 & FEROS \\
5729.6275  &   54.5  &3.9 & FEROS \\
5744.5979  &   68.0  &2.9 & FEROS \\
6047.6178  &   61.4  &2.8 & FEROS \\
6056.6058  &   60.7  &3.3 & FEROS \\
6066.6210  &   66.2  &2.8 & FEROS \\
6099.6025  &   59.7  &3.1 & FEROS \\
6110.5807  &   50.2  &2.7 & FEROS \\
6140.6110  &   37.0  &3.8 & FEROS \\
6321.7955  & -140.3  &2.8 & FEROS \\
6331.8191  & -133.8  &3.9 & FEROS \\
6342.7703  & -131.0  &3.8 & FEROS \\
6412.6435  & -162.3  &2.7 & FEROS \\
7072.8844  &  -20.5  &6.6 & FEROS \\
7388.8443  &  -11.1  &4.2 & FEROS \\
7471.9060  &    3.0  &3.8 & FEROS \\
7641.4875  &   18.2  &5.5 & FEROS \\
7012.8496  &   -17.5 &  4.2 & CHIRON \\
7050.7982  &   -19.4 &  4.2 & CHIRON \\
7079.7433  &   -12.8 &  4.4 & CHIRON \\
7101.6655  &     1.0 &  4.8 & CHIRON \\
7120.7170  &    -7.8 &  5.3 & CHIRON \\
7140.7324  &   -19.0 &  5.3 & CHIRON \\
7162.5652  &   -15.1 &  4.6 & CHIRON \\
7181.4933  &    -2.2 &  4.4 & CHIRON \\
7206.4763  &    -3.1 &  4.7 & CHIRON \\
7255.4986  &     3.2 &  4.6 & CHIRON \\
7260.4716  &   -15.3 &  5.2 & CHIRON \\
7270.5026  &    16.5 &  4.7 & CHIRON \\
7379.8748  &    15.0 &  4.3 & CHIRON \\
7391.8555  &     8.2 &  5.6 & CHIRON \\
7403.7968  &     2.3 &  5.5 & CHIRON \\
7404.8142  &    15.8 &  5.0 & CHIRON \\
7405.8539  &    18.2 &  5.4 & CHIRON \\
7491.6450  &    32.1 &  4.2 & CHIRON \\
\hline \vspace{-0.3cm} \\
\hline\hline
\end{tabular}
\end{table}

\end{appendix}

\end{document}